\documentclass[prl, superscriptaddress,showkeys, reprint]{revtex4-1}
\usepackage{titlesec}
\usepackage{graphicx}
\usepackage{dcolumn}
\usepackage{bm}
\usepackage{color}
\usepackage{tabularx}
\usepackage{tabularray}
\usepackage{multirow, makecell, booktabs}
\usepackage{threeparttable}
\usepackage{array}
\usepackage{amsmath}
\usepackage{amssymb}
\usepackage{mhchem}
\usepackage{comment}
\usepackage{ragged2e} 

\usepackage{pbox}
\usepackage{float}
\DeclareUnicodeCharacter{0394}{$\Delta$}

\usepackage{stmaryrd}  

\begin{document}

\preprint{APS/123-QED}

\title{Exciton-activated effective phonon magnetic moment in monolayer \ce{MoS2}}

\author{Chunli Tang}
\thanks{These two authors contributed equally}
\affiliation{Department of Electrical and Computer Engineering, Auburn University, Auburn, Alabama 36849, USA}
\affiliation{Department of Physics, Auburn University, Auburn, Alabama 36849, USA}

\author{Gaihua Ye}
\thanks{These two authors contributed equally}
\affiliation{Department of Electrical and Computer Engineering, Texas Tech University, Lubbock, Texas 79409, USA}

\author{Cynthia Nnokwe}
\affiliation{Department of Electrical and Computer Engineering, Texas Tech University, Lubbock, Texas 79409, USA}

\author{Mengqi Fang}
\affiliation{Department of Mechanical Engineering, Stevens Institute of Technology, Hoboken, New Jersey 07030, USA}

\author{Li Xiang}
\affiliation{National High Magnetic Field Laboratory, Tallahassee, Florida 32310, USA}

\author{Masoud Mahjouri-Samani}
\affiliation{Department of Electrical and Computer Engineering, Auburn University, Auburn, Alabama 36849, USA}

\author{Dmitry Smirnov}
\affiliation{National High Magnetic Field Laboratory, Tallahassee, Florida 32310, USA}

\author{Eui-Hyeok Yang}
\affiliation{Department of Mechanical Engineering, Stevens Institute of Technology, Hoboken, New Jersey 07030, USA}

\author{Tingting Wang}
\affiliation{Phonon Engineering Research Center of Jiangsu, School of Physics and Technology, Nanjing Normal University, Nanjing 210023, China}

\author{Lifa Zhang}
\affiliation{Phonon Engineering Research Center of Jiangsu, School of Physics and Technology, Nanjing Normal University, Nanjing 210023, China}

\author{Rui He}
\email{rui.he@ttu.edu}
\affiliation{Department of Electrical and Computer Engineering, Texas Tech University, Lubbock, Texas 79409, USA}

\author{Wencan Jin}
\email{wjin@auburn.edu}
\affiliation{Department of Physics, Auburn University, Auburn, Alabama 36849, USA}
\affiliation{Department of Electrical and Computer Engineering, Auburn University, Auburn, Alabama 36849, USA}

\date{\today}

\begin{abstract}
Optical excitation of chiral phonons plays a vital role in studying the phonon-driven magnetic phenomena in solids. Transition metal dichalcogenides host chiral phonons at high symmetry points of the Brillouin zone, providing an ideal platform to explore the interplay between chiral phonons and valley degree of freedom. Here, we investigate the helicity-resolved magneto-Raman response of monolayer \ce{MoS2} and identify a doubly degenerate Brillouin-zone-center chiral phonon mode at $\sim$270 $\mathrm{cm}^{-1}$. Our wavelength- and temperature-dependent measurements show that this chiral phonon is activated through the resonant excitation of $A$ exciton. Under an out-of-plane magnetic field, the chiral phonon exhibits giant Zeeman splitting, which corresponds to an effective magnetic moment of $\sim$2.5$\mu_B$. Moreover, we carry out theoretical calculations based on the morphic effects in nonmagnetic crystals, which reproduce the linear Zeeman splitting and Raman cross-section of the chiral phonon. Our study provides important insights into lifting the chiral phonon degeneracy in an achiral covalent material, paving a new route to excite and control chiral phonons. 

\end{abstract}

\maketitle
\section{I. Introduction}
\vspace{-10pt}
Circularly polarized phonons or chiral phonons are lattice vibration modes in which the atomic circular motions produce nonzero angular momentum~\cite{Zhang2014AngularEffect, Juraschek2019OrbitalPhonons}. As a result, chiral phonons carry orbital magnetic moments, serving as a fundamental element in realizing a wide variety of quantum phenomena such as the Einstein-de Hass effect ~\cite{Zhang2014AngularEffect}, phonon Hall effect~\cite{Sheng2006TheoryDielectrics,Li2020PhononTitanate},  ac Stark effect~\cite{Korenev2016Long-rangeStructure}, pseudogap phase of cuprates~\cite{Grissonnanche2020ChiralCuprates}, and interlayer exchange coupling in oxide heterostructures~\cite{Jeong2023UnconventionalHeterostructures}. 

Recent terahertz (THz) experiments have demonstrated coherent excitation of infrared-active phonon modes with controlled relative phases, giving rise to dynamical magnetization from lattice rotations in solids~\cite{Nova2017AnDrivenphonons}. One of the intriguing discoveries is that the chiral phonon magnetic moments can be 3 to 4 orders of magnitude greater than the theoretical calculation based on the Born effective charge and the ion masses~\cite{Juraschek2019OrbitalPhonons}. For example, the $E_u$ optical phonon in the Dirac semimetal \ce{Cd3As2} is excited by the circularly polarized THz laser pulses. The giant effective magnetic moment of $\sim$2.7$\mu_B$ ($\mu_B$ is electron Bohr magneton) is revealed by the phonon Zeeman effect, that is, the energy splitting between left-hand and right-hand phonon polarization in an external magnetic field~\cite{Cheng2020ASemimetal}. A similar size of phonon Zeeman splitting is also reported in narrow-gap semiconductor \ce{PbTe}~\cite{Baydin2022MagneticPbTe}, \ce{Pb_{1-x}Sn_xTe} thin films in the topological crystalline insulator phase ($\mathrm{x>0.32}$)~\cite{Hernandez2022ChiralInsulator}, and polar antiferromagnet \ce{Fe2Mo3O8}~\cite{wu2023fluctuation}. Remarkably, these studies all suggest electronic contributions in describing the magnetic moments of chiral phonons. In this regard, new mechanisms such as chiral phonon-induced electronic orbital magnetization~\cite{Xiong2022EffectivePhonons}, spin-chiral phonon coupling~\cite{Juraschek2022GiantParamagnets,Hamada2020ConversionRotation,Geilhufe2023ElectronmathrmKTaO_3,Luo2023LargeHalides}, and the topological contribution ~\cite{PhysRevLett.123.255901, Ren2021PhononMagnetization, PhysRevB.105.064303, PhysRevLett.130.226302} are proposed to explain dynamical magnetization. 

\begin{table*}[ht!]
\textbf{\caption{Valley selective chiral phonon phenomena in photoluminescence experiments in TMDC systems.}}
\vspace{5pt}
\centering
\renewcommand{\arraystretch}{1.7}
\begin{tabular}[c]{|>{\centering\arraybackslash}p{3.3cm}|>{\centering}p{2.2cm}|>{\centering}p{4cm}|>{\centering}p{3.3cm}|>{\centering}p{2.2cm}|>{\centering\arraybackslash}p{1.2cm}|}
\hline

\textbf{Phenomena}  & \textbf{Material}     &   \textbf{Optical excitation}   &  \textbf{Phonon mode}     & \textbf{g-factor}   & \textbf{Ref.}  \\
\hline
Infrared circular dichroism & \multirow{1.5}{2.2cm}{\centering CVD \ce{WSe2}}  & Dark A trion \\ Bright B trion  & LO($K$)=29$\pm$8 meV & N/A & \multirow{1.5}{1.2cm}{\centering~\cite{Zhu2018ObservationPhonons}}  \\
\hline

\multirow{8}{3.3cm}{\centering Chiral phonon replicas \\ in PL spectra} & \multirow{8}{*}{ $\ce{BN/WSe{2}/BN}$} & \multirow{1.5}{4cm}{\centering Intravelly dark exciton} & \multirow{5}{*}{ $E^{''}(\Gamma)$\newline $\sim$21.4-21.6 meV} & -9.4$\pm$0.1 \\ -9.9$\pm$0.1 & \multirow{1.5}{1.2cm}{\centering~\cite{Li2019EmergingWSe2,Liu2019Valley-selectivemathrmWSmathrme_2}}\\ \cline{5-6}\cline{3-3}

& & Intravalley positive \\ dark trion & & \multirow{1.5}{2.2cm}{\centering-9.9$\pm$0.2} & \multirow{3}{1.2cm}{\centering~\cite{Liu2019Valley-selectivemathrmWSmathrme_2}}\\ \cline{5-5}\cline{3-3}
& & Intravalley negative \\ dark trion & & \multirow{1.5}{2.2cm}{\centering-9.2$\pm$0.1} & \\ \cline{3-6}
& & Intervalley dark exciton & $E^{''}(K)\sim$26.6 meV & 12.5 & \multirow{3}{*}{~\cite{Li2019Momentum-DarkPhonon,Liu2020MultipathmathrmWSe_2,He2020ValleySemiconductor}}\\ \cline{3-5}
& & Intravelly positive  exciton & $E^{''}(\Gamma)\sim$24.1 meV & 13.7 & \\ \cline{3-5}
& & Intravelly negative  exciton & $E^{''}(K)\sim$26.6 meV & 11.9 & \\ \cline{3-5}
\hline

\multirow{3}{3.3cm}{\centering Magnetophonon resonance at $\sim$24 T} & \multirow{3}{2.2cm}{\centering $\ce{MoSe{2}/WSe2}$ heterobilayer} & \multirow{3}{2.8cm}{\centering Interlayer exciton} & $E^{''}(\Gamma)$ of $\ce{MoSe2}$ and $\ce{WSe2}$ & \multirow{1.5}{2.2cm}{\centering-16 (H-type)} & \multirow{1.5}{1.2cm}{\centering~\cite{Delhomme2020FlippingHeterobilayers}} \\ \cline{4-6}

& & & \multirow{1.5}{3.3cm}{\centering TA phonons of $\ce{MoSe2}$} & -14.8 (H-type)\newline+4.7 (R-type) & \multirow{1.5}{1.2cm}{\centering~\cite{Smirnov2022Valley-magnetophononExcitons}} \\
\hline

Doublets of quantum dot like peaks in PL & \multirow{1.5}{2.2cm}{\centering$\ce{WSe2}$/BN} & Quantum dot emission & \multirow{1.5}{3.3cm}{\centering $E^{''}(\Gamma)\sim$21.8 meV} & \centering-8.9$\pm$1.1\\-9.3$\pm1.9$ & ~\cite{Chen2019EntanglementWSe2} \\

\hline
\end{tabular}

\begin{tablenotes}
    \item \textbf{Note}: The g-factors of the phonon replicas in this table reflect the values of their corresponding excitonic states, rather than the intrinsic phonon magnetic moment.
\end{tablenotes}
  
\end{table*}
\label{table:Tabel1}

In a broader sense, degenerate chiral phonons that consist of superpositions of two orthogonal phonon components can be resonantly excited with light~\cite{Chen2019ChiralMaterials}. This mechanism enables the chirality-governed emission channels of phonons and photons, which have been implemented through the valley degree of freedom in monolayer transition metal dichalcogenides (TMDCs). The threefold rotational symmetry and spatial inversion symmetry breaking in monolayer TMDCs can endow chiral phonons at high symmetry points of the Brillouin zone (BZ)~\cite{Zhang2015ChiralLattices}. Since the original discovery of chiral phonon at the BZ corner of \ce{WSe2}~\cite{Zhu2018ObservationPhonons}, various valley-selective chiral phonon phenomena (summarized in \textbf{Table I}) have been identified in magneto-photoluminescence (PL) spectra involving dark excitons~\cite{Li2019EmergingWSe2,Liu2019Valley-selectivemathrmWSmathrme_2,Li2019Momentum-DarkPhonon}, trions~\cite{Zhu2018ObservationPhonons,Liu2019Valley-selectivemathrmWSmathrme_2,Liu2020MultipathmathrmWSe_2,He2020ValleySemiconductor}, interlayer excitons ~\cite{Delhomme2020FlippingHeterobilayers,Smirnov2022Valley-magnetophononExcitons}, and quantum dot emission ~\cite{Chen2019EntanglementWSe2} in TMDC systems. These observations leave many open questions concerning the connection between the excitonic states and chiral phonons, for example: Aside from dark excitons, trions, and interlayer excitons, can chiral phonons be generated through the decay of bright excitons? While the g-factors of the phonon replicas in PL spectra only ubiquitously reflect the valley Zeeman splitting of the original excitonic states, what is the intrinsic chiral phonon magnetic moment in TMDCs? In an achiral and nonmagnetic material, how can one lift the degeneracy of the chiral phonons and implement functional optical activity? Addressing these questions is crucial for understanding the magneto-optical properties based on chiral interactions. However, direct access to excitons and chiral phonons simultaneously with helicity resolution has not been demonstrated.

Helicity-resolved Raman spectroscopy (HRRS) has been proven to be a powerful tool for assigning phonon symmetry in TMDCs~\cite{Chen2015Helicity-ResolvedLayers,Zhao2022TheMaterials,Lacinska2022Raman1T-TaS2}. In the off-resonance condition, the helicity selection rule shows that the helicity-conserved $A_{1g}$/$A_{1}^{'}$ modes can be detected in the $\sigma^{+} \sigma^{+}$ and $\sigma^- \sigma^-$ channels, while the helicity-changed $E_{2g}$/$E^{'}$ modes can be detected in the $\sigma^+ \sigma^-$ and $\sigma^- \sigma^+$ channels, where $\sigma^i \sigma^s$  ($i,s = +,-$) denotes helicities of the incident and scattered light~\cite{Tatsumi2018ConservationScattering}. The on-resonance excitation leads to modified Raman tensors in the helicity selection rule, revealing the nature of exciton-phonon coupling from the deformation potential or the Fr\"{o}hlich interaction~\cite{Zhao2020CharacterizationLight}. Remarkably, recent studies show that HRRS is sensitive to chirality-related phenomena such as chiral phonon splitting in \ce{\alpha-HgS}~\cite{Ishito2023Truly-HgS}, chiral charge density wave~\cite{Yang2022VisualizationWaves}, and ferro-rotational domain states in \ce{1T-TaS2}~\cite{Liu2023ElectricalCrystals}. In this work, using HRRS on resonance with the $A$ exciton of monolayer \ce{MoS2}, we identify a new broad mode at $\sim$270 cm$^{-1}$ and the selection rule analysis demonstrates that this mode is a doubly degenerate chiral phonon ($\Omega_{\pm}$). By applying an out-of-plane magnetic field, we break both the energy and intensity degeneracy of the chiral phonon. The phonon Zeeman splitting in the $\sigma^+ \sigma^-$ and $\sigma^- \sigma^+$ channels reveals an effective phonon magnetic moment of $\sim2.5\mu_B$, which is approximately 6 orders of magnitude greater than the theoretically predicted value of TMDCs. Moreover, the magnetic field can control the helicity polarization ($\rho$) of the $\Omega_{+}$  and $\Omega_{-}$ modes in a linear fashion up to 50$\%$ at $\pm$7 T. In contrast to the behavior of thermally populated phonons, our temperature-dependent measurements show that this chiral phonon appears at $\sim$180 K and its spectral intensity grows rapidly with descending temperature. The onset of the chiral phonon is concurrent with the enhanced PL intensity of $A$ exciton, indicating this chiral phonon is activated through the bright $A$ exciton transition. Finally, we carry out theoretical calculations based on the Morphic effects in nonmagnetic crystals, which reproduce the linear Zeeman splitting and Raman cross-section of the chiral phonon. 

\begin{figure*}[ht!]
\includegraphics[width=0.98\textwidth]{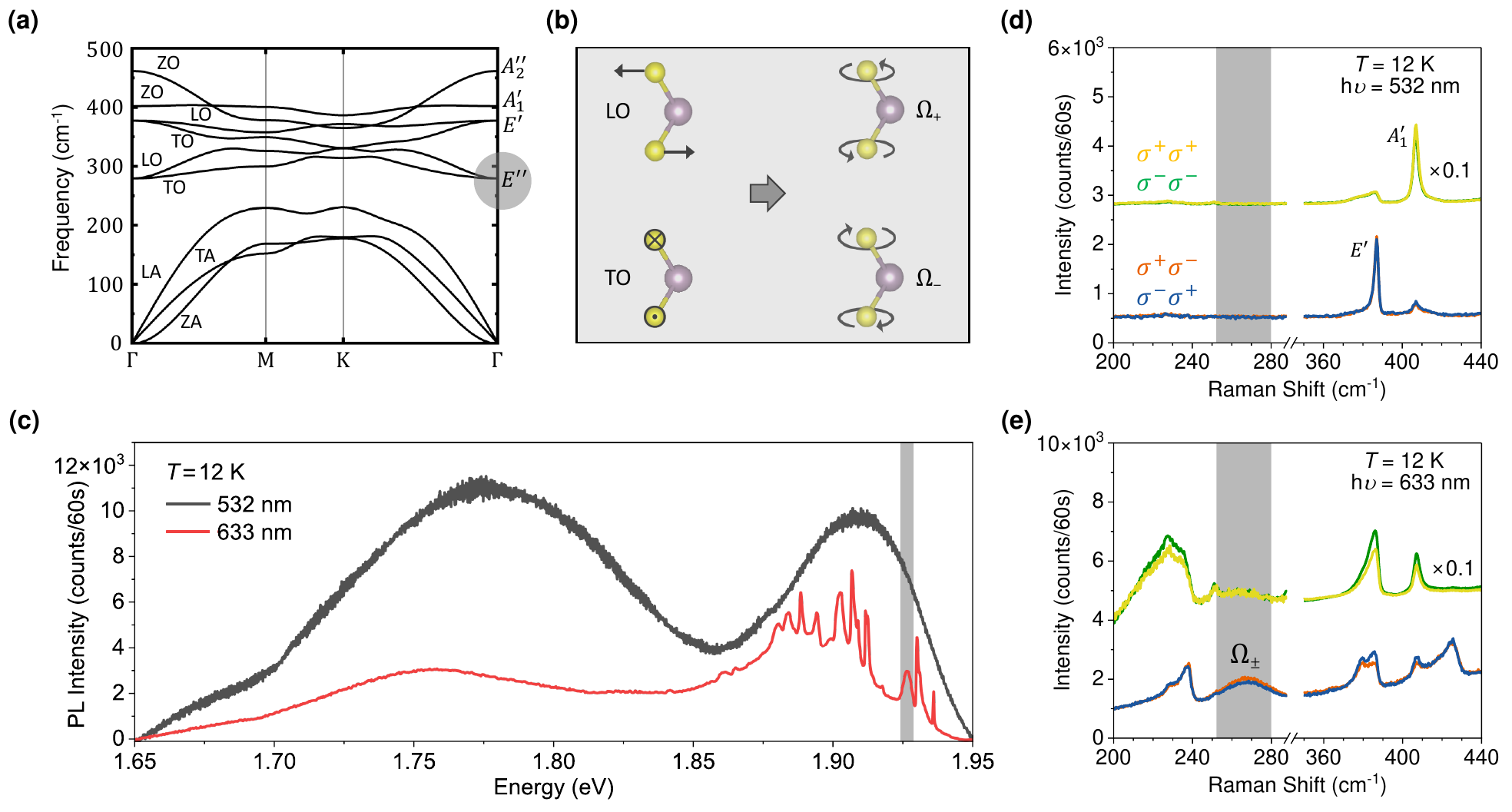}
\caption{\label{SelectionRule} \textbf{Helicity selection rule of the doubly degenerate chiral phonon in monolayer MoS$_2$.} \textbf{(a)} Calculated phonon dispersion of monolayer MoS$_2$. \textbf{(b)} Schematics of the doubly degenerate BZ center E$^{''}$ phonon modes. The superposition of the LO and TO modes gives rise to the left- and right-handed chiral phonon modes ($\Omega_\pm$). \textbf{(c)} PL spectra acquired in the $\sigma^+\sigma^-$ channel at 12 K using 532 nm (2.33 eV) and 633 nm (1.96 eV) excitation, respectively. Helicity selection rule of the Raman modes in the \textbf{(d)} off-resonance and \textbf{(e)} on-resonance condition in the $\sigma^+ \sigma^+$ (yellow), $\sigma^- \sigma^-$ (green), $\sigma^+\sigma^-$ (orange), and $\sigma^- \sigma^+$ (blue) polarization channels. The grey shaded areas in \textbf{(c)-(e)} highlight the chiral phonon modes. The spectra in the range of 340-440 $\mathrm{cm}^{-1}$ in the $\sigma^+ \sigma^+$ and $\sigma^- \sigma^-$ channels are scaled by a factor of 0.1.}
\end{figure*}

\section{II. Experimental Results}
\vspace{-10pt}
\subsection{A. Helicity selection rule}
\vspace{-10pt}
The monolayers \ce{MoS2} are synthesized on a \ce{SiO2}/Si substrate via chemical vapor deposition (CVD)~\cite{Fu2020EnablingIron-doping,Wang2017Location-specificSize}. Atomic force microscope characterization of a \ce{MoS2} flake is shown in Fig. S1 in the Supplementary Material~\cite{Supplementary}. We first compare the PL spectra acquired at 12 K using 532 nm (2.33 eV) and 633 nm (1.96 eV) excitation (See Fig. \ref{SelectionRule}(c)). While the $A$ exciton ($\sim$1.90 eV) and broad localized emitter peak ($\sim$1.65-1.85 eV) are found in the PL spectrum at 532 nm excitation, a series of Raman peaks appear in the 633 nm spectrum. By comparing these Raman peaks with the existing resonant Raman spectrum of \ce{MoS2} (see Fig. S2 in the Supplementary Material~\cite{Supplementary}), we identify a new mode at $\sim$270 cm$^{-1}$ (absolute energy of $\sim$1.927 eV as marked by the grey shade in Fig. \ref{SelectionRule}(c)), which matches the energy of the BZ center $E^{''}$ phonon mode as highlighted in the calculated phonon dispersion of monolayer \ce{MoS2} in Fig. \ref{SelectionRule}(a). As illustrated in Fig. \ref{SelectionRule}(b), the superposition of two orthogonal linear vibrations (LO and TO modes) results in doubly degenerate chiral phonons denoted as $\Omega_\pm$. We note that a sharp Raman mode at 287 cm$^{-1}$ was previously reported in $2H$-\ce{MoS2} and Raman mapping demonstrates that it is an $E_{1g}$ mode that is only detectable at the edges~\cite{Guo2023DiscoveringMaterials}. Even though Raman mapping at low temperature is not feasible for our experimental setup, we have carefully surveyed the body and edge regions on our $\ce{MoS2}$ flake and confirmed that the chiral phonon mode is not an edge mode.

To verify the chiral nature of the new mode, we carry out helicity-resolved Raman measurements. In the off-resonance condition (532 nm, see Fig. \ref{SelectionRule}(d)), the in-plane $E^{'}$ mode appears at 384.3 cm$^{-1}$ in $\sigma^+ \sigma^-$ and $\sigma^- \sigma^+$ channels, while the out-of-plane $A_{1}^{'}$ mode appears at 405.4 cm$^{-1}$ in the $\sigma^+ \sigma^+$ and $\sigma^- \sigma^-$ channels, which agrees with the HRRS results in the literature~\cite{Chen2015Helicity-ResolvedLayers, PhysRevB.95.165417}. Also, the $\sim$21 cm$^{-1}$ frequency difference between $E^{'}$ and $A_{1}^{'}$ modes is consistent with the CVD-grown monolayer \ce{MoS2}~\cite{Zhang2015PhononMaterial}.

\begin{figure*}[ht!]
\includegraphics[width=0.98\textwidth]{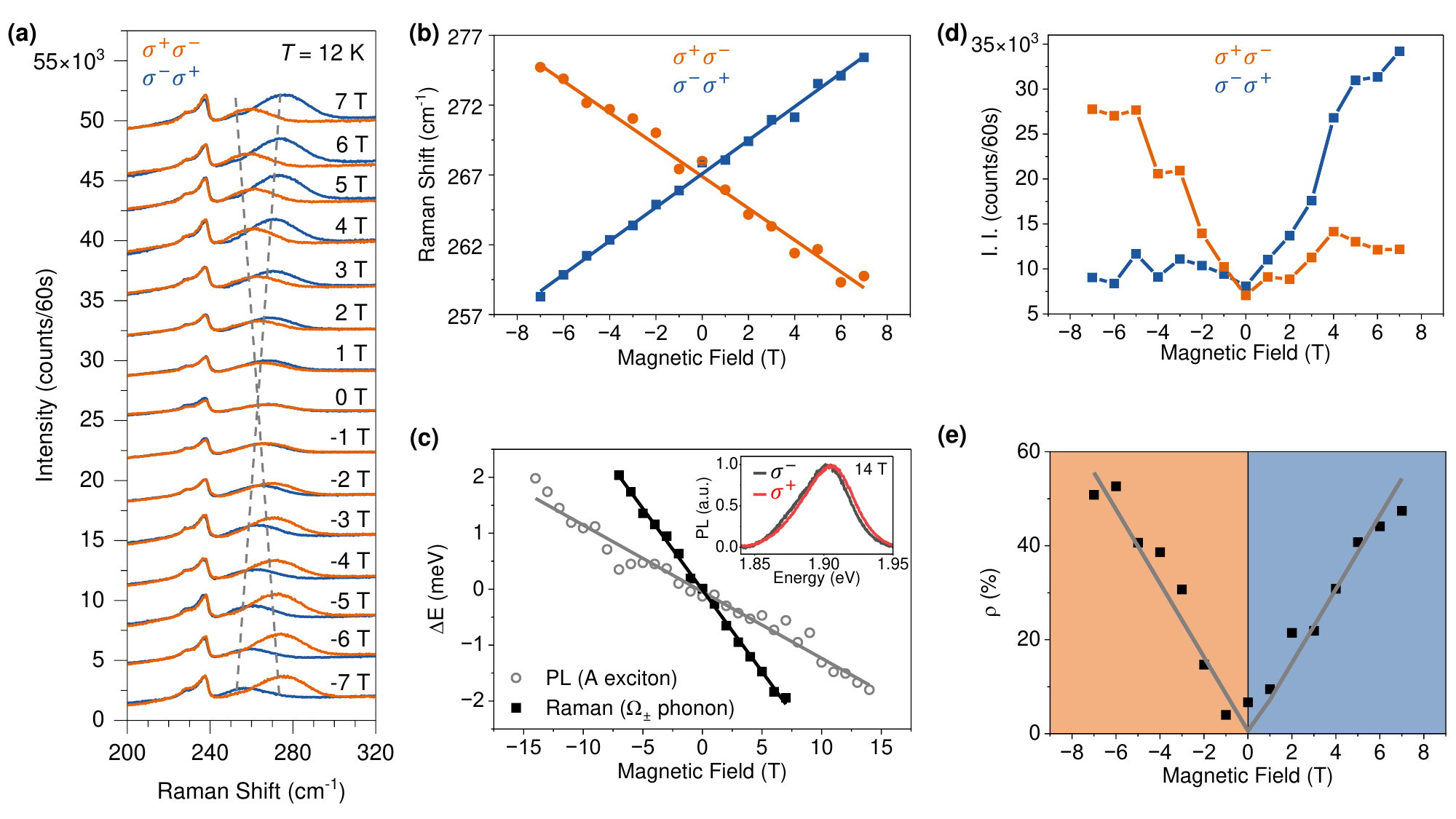}
\caption{\label{Field} \textbf{Broken frequency and intensity degeneracy of chiral phonon in the magnetic field. (a)} Magnetic field-dependent Raman spectra in the $\sigma^+ \sigma^-$ (orange) and $\sigma^- \sigma^+$ (blue) polarization channels at the excitation wavelength of 633 nm at 12 K. The spectra are vertically offset for clarity. The grey dashed lines are the guide to the eye of the phonon Zeeman splitting. \textbf{(b)} Split chiral phonon frequencies as a function of the magnetic field with linear fits. \textbf{(c)} Comparison between the valley Zeeman splitting of the $A$ exciton determined by magneto-PL (grey circles) and the chiral phonon Zeeman splitting determined by Raman (black squares). Inset shows the split $\sigma^{+}$ and $\sigma^{-}$ PL peaks at 14 T. \textbf{(d)} Integrated intensity (I.I.) of the split chiral phonon as a function of the magnetic field. \textbf{(e)} Helicity polarization ($\rho$) as a function of the magnetic field with the linear fit. The blue and orange shades indicate the dominant Raman signal.}
\end{figure*}

Under the resonant excitation at 633 nm, the breakdown of the helicity selection rule occurs for $E^{'}$ and $A_{1}^{'}$ modes due to the exciton-phonon coupling~\cite{Zhao2020CharacterizationLight, PhysRevB.95.165417}. For the new mode at $\sim$270 cm$^{-1}$, as shown in Fig. \ref{SelectionRule}(e), it only appears in the helicity-changed $\sigma^+ \sigma^-$ and $\sigma^- \sigma^+$ channels with identical frequency and intensity, and is absent in the helicity-reserved $\sigma^+ \sigma^+$ and $\sigma^- \sigma^-$ channels. We can thus retrieve the Raman tensor of this mode in the form of $\mathrm{\textbf{$R$} = \begin{pmatrix}
a & c \\
c & -a 
\end{pmatrix}}$
or 
$\mathrm{\textbf{$R^{'}$} = \begin{pmatrix}
a & -c \\
-c & -a 
\end{pmatrix}}$. The nonzero off-diagonal term $c$ guarantees that $R$ and $R^{'}$ are connected by a mirror operation, representing Raman modes with opposite chirality. Therefore, we assign the new mode as the doubly degenerate chiral phonon ($\Omega_{\pm}$). The resonant excitation plays a key role in the activation of the chiral phonon because the $E^{''}$ mode is strictly undetectable in the backscattering geometry under the off-resonance condition (see Fig. 1(d)). Moreover, we examine the selection rule in the linear polarization channels (see Fig. 5 in APPENDIX B). The linearly parallel and crossed channels have the same Raman intensity and exhibit no angular anisotropy upon in-plane rotation of the sample. We thus determine the diagonal term $c=ai$. The detailed derivation of the Raman tensors can be found in APPENDIX B. The complex Raman tensor and the relative phase between the diagonal and off-diagonal tensor elements indicate that the resonant excitation and anisotropic electron-phonon interaction may play a key role in the activation of this chiral phonon ~\cite{kim2021polarized, han2022complex, resende2020origin, pimenta2021polarized}. It is worth noting that an axial Higgs mode in the charge density wave system \ce{RTe3} was recently detected by quantum interference between the pathways with symmetric and antisymmetric contributions to the Raman tensor ~\cite{wang2022axial}. Our findings of the complex Raman tensor provide a new method to detect axial/chiral modes using resonant scattering.
\vspace{-10pt}

\subsection{B. Phonon Zeeman splitting}
\vspace{-10pt}
To quantify the angular momentum of the chiral phonon mode, we carry out helicity-resolved magneto-Raman measurements. Figure \ref{Field}(a) shows the Raman spectra in the $\sigma^+ \sigma^-$ and $\sigma^- \sigma^+$ channels acquired at 12 K with varying magnetic fields from -7 T to 7 T. The application of a magnetic field lifts the degeneracy of the chiral phonon, leading to the split $\Omega_{+}$ and $\Omega_{-}$ components in the $\sigma^+ \sigma^-$ and $\sigma^- \sigma^+$ channels, respectively. As shown in Fig. \ref{Field}(b), the frequency of the split modes exhibits a linear response as a function of the magnetic field. Fitting the experimental data to the Zeeman splitting $\mathrm{\Omega_{\pm} = \Omega_{0}\pm\gamma B}$ yields the slopes of $\mathrm{1.14\pm0.03}$ cm$^{-1}$/T and $\mathrm{1.20\pm0.02}$ cm$^{-1}$/T, corresponding to $\mathrm{(2.44\pm0.06)\mu_{B}}$ and $\mathrm{(2.57\pm0.04)\mu_B}$, respectively. In the magneto-PL measurements summarized in \textbf{Table I}, the chiral phonon replicas and their corresponding excitonic states have the same g-factors~\cite{Li2019EmergingWSe2,Liu2019Valley-selectivemathrmWSmathrme_2,Li2019Momentum-DarkPhonon,Liu2020MultipathmathrmWSe_2,He2020ValleySemiconductor,Delhomme2020FlippingHeterobilayers,Smirnov2022Valley-magnetophononExcitons,Chen2019EntanglementWSe2}, only reflecting the properties of valley Zeeman splitting. In Fig.~\ref{Field}(c), we compare the chiral phonon Zeeman splitting ($\Omega_{\pm}$) and valley Zeeman splitting of $A$ exciton and find they are different in our CVD sample. The $\sim$2.5$\mu_{B}$ effective phonon magnetic moment of $\ce{MoS2}$ is comparable with the value of $\ce{Cd3As2}$~\cite{Cheng2020ASemimetal}. In this context, the relative splitting $\mathrm{\Delta\Omega/\Omega_{0}}$ in an external magnetic field of 50 T is expected to be $\sim$0.4, which is approximately 6 orders of magnitude larger than the predicted values for TMDCs~\cite{Juraschek2019OrbitalPhonons}.

In addition to frequency splitting, the $\Omega_{+}$ and $\Omega_{-}$ components in the $\sigma^+ \sigma^-$ and $\sigma^- \sigma^+$ channels have distinct Raman intensities. Figure \ref{Field}(d) shows the integrated Raman intensity of $\Omega_{+}$ and $\Omega_{-}$ components as a function of the magnetic field. In the positive magnetic field, Raman scattering is dominated by the chiral phonon component in the $\sigma^{-}\sigma^{+}$ channel. By flipping the magnetic field, Raman scattering from the opposite component becomes dominant. The different scattering intensities from opposite phonon chirality can be characterized by the helicity polarization defined as 
\begin{equation}
    \mathrm{\rho=\left| \frac{I(\sigma^{+} \sigma^{-})-I(\sigma^{-} \sigma^{+})}{I(\sigma^{+} \sigma^{-})+I(\sigma^{-} \sigma^{+})} \right| \times100\%}
\end{equation}
\noindent As shown in Fig. \ref{Field}(e), the helicity polarization evolves linearly with the magnetic field, arriving at 50$\%$ at $\pm$7 T. It is worth noting that the frequency and intensity of the $E^{'}$ mode (384.3 cm$^{-1}$) remain unchanged under magnetic field (see data in Supplementary Material Section S3~\cite{Supplementary}). This is mainly because the vibration of $E^{''}$ mode breaks the horizontal mirror symmetry of \ce{MoS2} plane, allowing strong coupling with spin; while spin-phonon coupling is essentially negligible for $E^{'}$ mode ~\cite{shin2018phonon, zhang2023light}. Recently, HRRS measurements of van der Waals magnets $\ce{Fe3GeTe2}$ (D$_{6h}$ point group)~\cite{Du2019LatticeFe3GeTe2} and $\ce{CrBr3}$ (S$_6$ point group)~\cite{Yin2021ChiralMagnet} identify $E_{2g}/E_{g}$ phonon modes in the helicity-changed $\sigma^{+}\sigma^{-}$ and $\sigma^{-}\sigma^{+}$ channels. However, these modes display no magnetic field dependence. Therefore, we argue that observation of doubly degenerate $E_{2g}/E_{g}/E^{'}$ phonon modes that reverse the helicity of incidence photon cannot guarantee the nontrivial chirality-dependent magneto-optical properties. The coupling between the chiral phonons and the spin-valley degree of freedom must be considered as the key ingredient.

\begin{figure}
\includegraphics[width=0.48\textwidth]{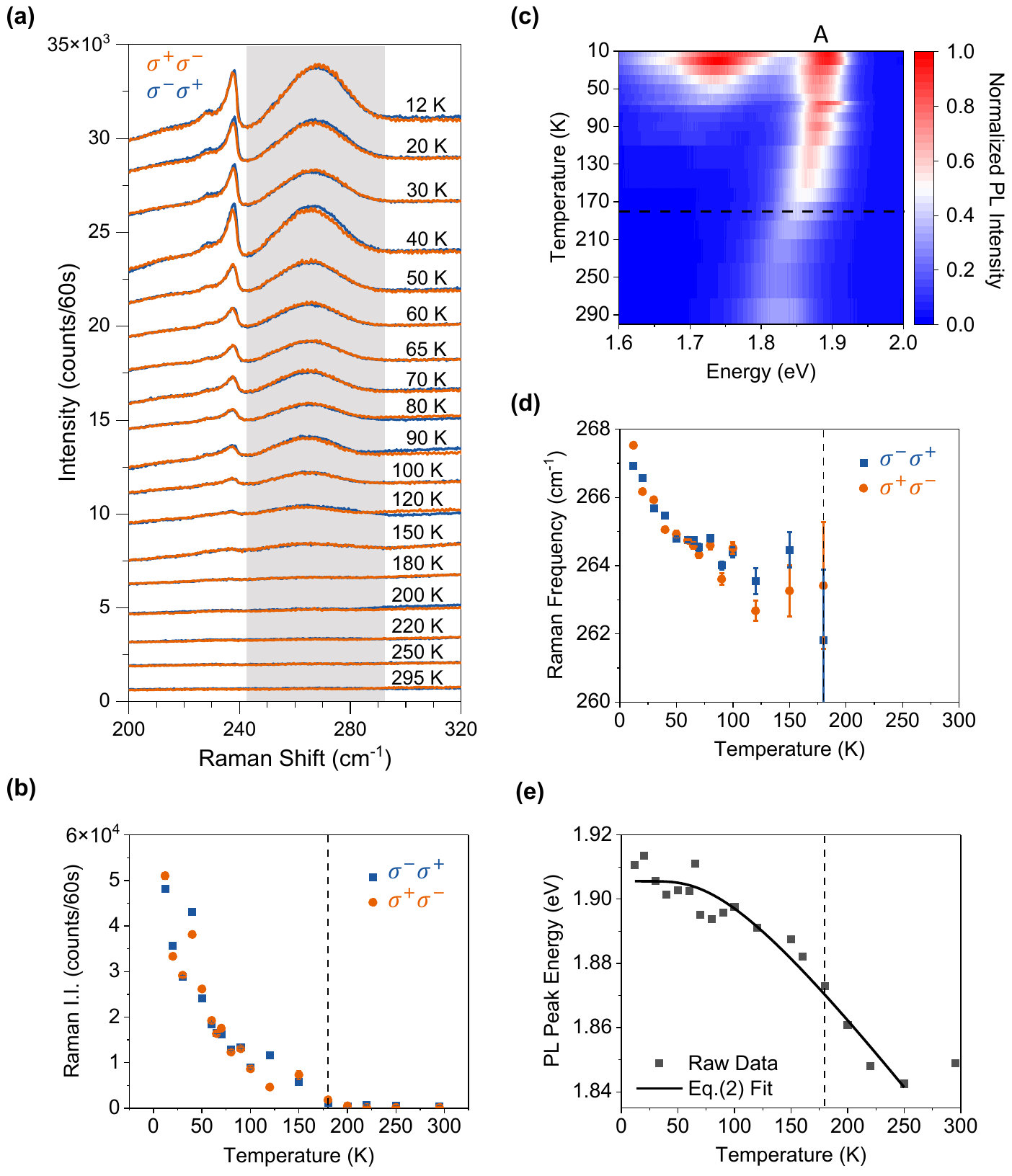}
\caption{\label{Temperature} \textbf{Chiral phonon activated through exciton transition. (a)} Temperature-dependent Raman spectra in the $\sigma^+ \sigma^-$ (orange), and $\sigma^- \sigma^+$ (blue) polarization channels at the excitation wavelength of 633 nm. The spectra are vertically offset for clarity. \textbf{(b)} Integrated intensity (I.I.) of the chiral phonon mode as a function of temperature. \textbf{(c)} Heatmap of the temperature-dependent PL spectra measured using 532 nm excitation laser. “A” denotes the $A$ exciton emission. The colorscale represents the normalized PL intensity. \textbf{(d)} Raman frequency of the $\Omega_{\pm}$ mode and \textbf{(e)} PL peak energy of the $A$ exciton as a function of temperature. The solid black curve in (e) is the fit to Eq. (1). The black dashed lines in panels (b)-(e) mark the temperature of 180 K where both Raman ($\Omega_{\pm}$) and PL ($A$ exciton) spectral intensities rapidly grow.}
\end{figure}

We also conduct HRRS measurements in pristine exfoliated monolayer \ce{MoS2} flakes and reproduce the selection rule and phonon Zeeman splitting we obtained from the CVD samples (see Supplementary Material Section S4~\cite{Supplementary}). However, the profound background from the tail of the photoluminescence in exfoliated samples leads to the skewed lineshape of the chiral phonon, hindering further analysis of the helicity polarization. Since the low intensity of the background in the CVD samples allows more consistent analysis of the chiral phonon, the data we show in this work are from the CVD samples unless otherwise specified. Moreover, we examine the laser power dependence of the split modes at 7 T and find that their integrated intensities linearly grow with the increasing laser power, which is consistent with the $E^{'}$ and $A_1^{'}$ phonon modes (see Supplementary Material Section S5~\cite{Supplementary}). 

\subsection{C. Temperature-dependent measurements}
\vspace{-10pt}
To further investigate the origin of this chiral phonon mode, we carry out temperature-dependent HRRS measurements. As shown in Fig. \ref{Temperature}(a), the chiral phonon mode emerges at $\sim$180 K and its integrated intensity (I.I.) shown in Fig. \ref{Temperature}(b) rapidly grows with decreasing temperature. In parallel, Fig. \ref{Temperature}(c) shows the temperature evolution of the PL spectra. As the broad localized emitter peak vanishes above 80 K, the $A$ exciton dominates the PL spectra up to 295 K and the enhanced $A$ exciton emission below $\sim$180 K is due to the quenched non-radiative recombination at low temperature. The emergence of the chiral phonon mode is concurrent with the enhanced PL intensity of $A$ exciton at $\sim$180 K, supporting the picture of exciton-activated chiral phonon. Although $A$ exciton and the chiral phonon mode are intimately connected, they exhibit distinct energy shifts in the temperature-dependent measurements. As shown in Fig. \ref{Temperature}(d), the chiral phonon mode shifts by $\sim$5 cm$^{-1}$ ($\sim$0.6 meV) upon cooling from 180 K to 12 K. In contrast, the $A$ exciton shifts toward higher energy by $\sim$35 meV in this temperature range (see Fig. \ref{Temperature}(e)). Similar results are obtained in the exfoliated monolayer \ce{MoS2} (see Fig. S6 in the Supplementary Materials ~\cite{Supplementary}).

We also directly compare the absolute photon energies of the chiral phonon mode and the $A$ exciton through the temperature-dependent PL spectra at 633 nm excitation and confirm that they follow different trends (See Supplementary Materials Section S7~\cite{Supplementary}). The shift of $A$ exciton can be attributed to the change of the semiconducting bandgap, which can be fitted to an equation proposed in Ref.~\cite{ODonnell1991TemperatureGaps}
\begin{equation}
    E_{g}(\emph{T})=E_{0} -S\left \langle \hbar\omega \right \rangle[\mathrm{coth}(\frac{\left\langle \hbar\omega \right \rangle}{k_{B}T})-1]
\end{equation}
\noindent where $\mathrm{\left \langle \hbar\omega \right \rangle}$ and $\emph{S}$ are the average phonon energy and the coupling parameter, respectively. Remarkably, the rapid shift of the Raman frequency without saturation (Fig. \ref{Temperature}(d)) is distinct from the temperature-dependent behavior of a typical phonon, indicating complicated mechanism underlying this mode. First, the broad linewidth of this mode suggests that it may be strongly impacted by disorders. Second, the temperature dependence of the Raman frequency is reminiscent of the anharmonic phonon in SnSe ~\cite{PhysRevB.98.224309}. Considering phonon Hamiltonian with lowest-order anharmonicity, the phonon stiffens monotonically with the descending temperature (See Supplementary Materials Section S7 ~\cite{Supplementary}), which is consistent with the trend of our experimental data. Despite these plausible interpretations, the origin and properties of this chiral phonon mode deserve further experimental and theoretical investigations.

\begin{figure}[t!]
\includegraphics[width=0.5\textwidth]{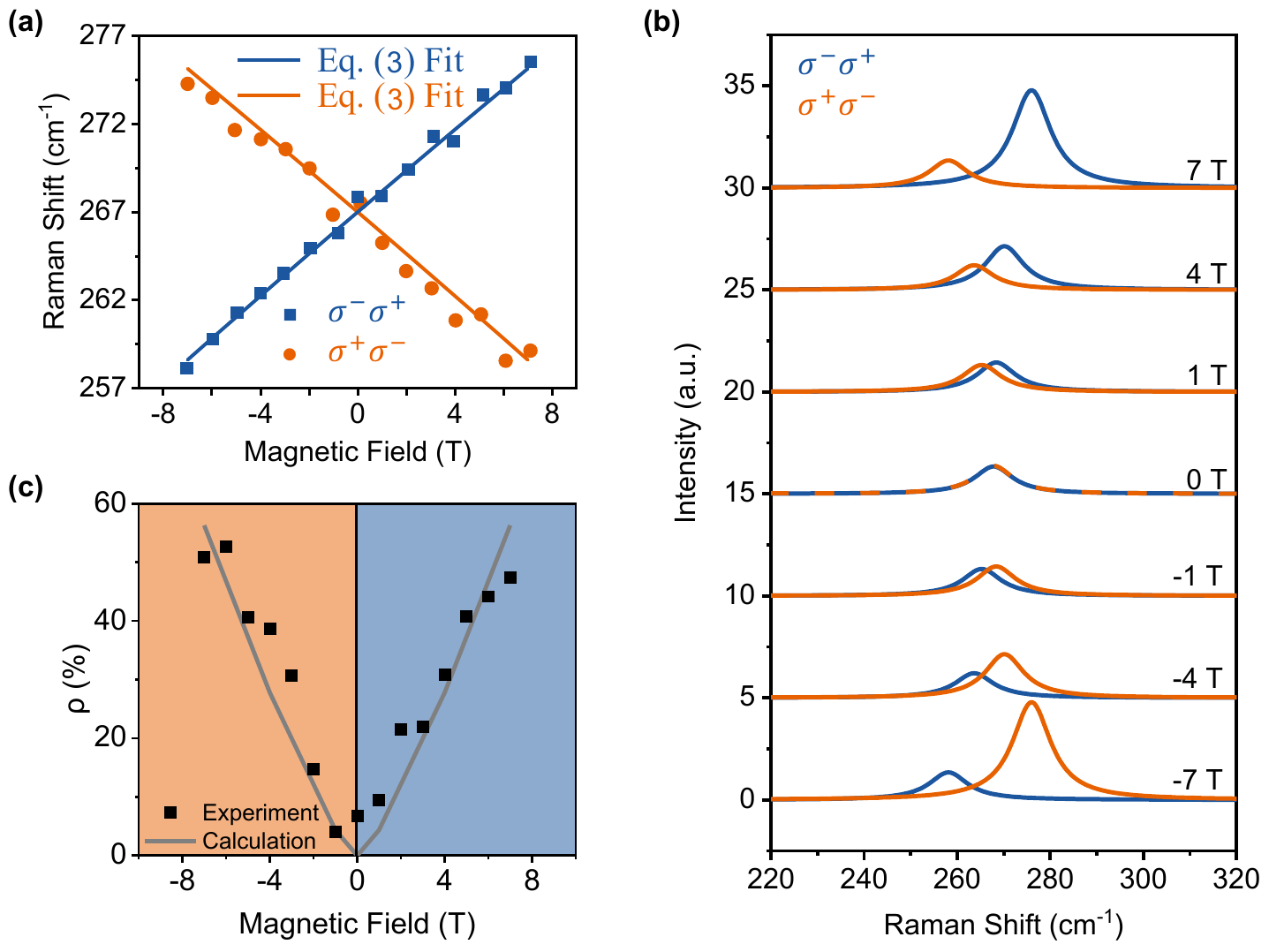}
\caption{\label{DFT} \textbf{Theoretical calculations of chiral phonon activities in magnetic field. (a)} Fitting of the split chiral phonon frequencies to Eq. (3) derived from the morphic effects. \textbf{(b)} Calculated Raman scattering cross-section in the $\sigma^{+}\sigma^{-}$ (orange) and $\sigma^{-}\sigma^{+}$ (blue) polarization channels at varying magnetic field. \textbf{(c)} Helicity polarization of Raman scattering cross-section as a function of magnetic field exacted from (b). The experimental data (solid square) are overlaid for comparison.}
\end{figure}

\section{III. Theoretical calculations}
\vspace{-10pt}
To evaluate our experimental results, we carry out theoretical calculations based on the morphic effects ~\cite{anastassakis1971morphic2,anastassakis1972morphic3,anastassakis1972morphic4}, a phenomenological treatment of the impact of an external magnetic field on the Raman scattering of optical phonons in nonmagnetic crystals ~\cite{schaack1976observation, schaack1977magnetic}. The large effective magnetic moment of the chiral phonon in \ce{MoS2} may be due to the interplay between the chiral phonon, exciton, and the specific band structure of \ce{MoS2} ~\cite{PhysRevLett.111.026601, PhysRevLett.115.176801}, while a microscopic theory is beyond the scope of this experimental work.

The detailed calculation methods can be found in APPENDIX C. The presence of an external static magnetic field ($\textbf{\emph{H}}=H e_z$) introduces an additional term in the effective spring constant $K_j(H)=K_j+K_{jH}$, where $(K_{jH})_{\alpha\beta}=(\frac{\partial K_{j\alpha\beta}}{\partial H_{\gamma}})H_{\gamma}$ is defined for the doubly degenerate phonon mode $E^{''}$ with a frequency of $\omega_j$. Symmetry analysis shows that $K_jH$ is antisymmetric, allowing non-zero $K_{12}=-K_{21}=K$. Thus, we obtain

\begin{equation}
\omega^{\pm}_{jH} = \sqrt{\omega^2_j\pm KH} \approx \omega_j \pm \frac{K}{2\omega_j}H
\end{equation} 

\noindent and the corresponding polarization vectors $\Tilde{\epsilon}_{j\pm} = \frac{1}{\sqrt{2}}(\epsilon_{j1},\pm i\epsilon_{j2})$ are circularly polarized, corresponding to the chiral phonon eigenmodes. This model can explain the linear Zeeman splitting of the $E^{''}$ phonon in \ce{MoS2}, in contrast to the quadratic magnetic field dependence of the TO phonon in PbTe ~\cite{Baydin2022MagneticPbTe}.

By fitting this model to our experimental data (see Fig.\ref{DFT}(a)), we obtain $K\simeq625$ $\mathrm{cm^{-2}/T}$. We then simulate the HRRS process by calculating the susceptibility tensor using the phonon eigenmodes. Based on the Raman scattering cross-section $I\propto |\langle e_{s} | \frac{\partial \chi }{\partial q_{i}} |e_{i}\rangle|^{2}(n_{i}+1)$, we implement the helicity selection rule and produce the magnetic field-dependent HRRS spectra in Fig. \ref{DFT}(b). The corresponding helicity polarization (see Fig. $\ref{DFT}$(c)) agrees with the experimental data very well.

\section{IV. Conclusion}
\vspace{-10pt}
In summary, our work features the first direct observation of chiral phonon derived from the doubly degenerate $E^{''}$ mode in monolayer \ce{MoS2} using helicity-resolved Raman scattering. We demonstrate the resonant excitation of $A$ exciton activates the chiral phonon and produces the helicity selection rule. The chiral phonon exhibits large effective magnetic moment and distinct Raman scattering intensity from the opposite chirality, which holds promise for phonon transport and quantum information applications. Our findings can be extended to other two-dimensional hexagonal lattices and shed light on manipulating chiral phonons through the magneto-optical effect.\\

\begin{center}
    \textbf{APPENDIX A: Experimental Methods}
\end{center}

Raman spectra were taken in a backscattering geometry using a Horiba LabRAM HR Evolution Raman Microscope system with a thermoelectrically cooled multichannel detector (CCD). The spectrometer is equipped with half- and quarter-wave plates and polarizers for linear and circular polarization studies. A 40$\times$ objective lens was used. The power of both 532 nm and 633 nm lasers was kept below 0.5 mW. The Raman and photoluminescence data were acquired using 1800 grooves/mm grating and 600 grooves/mm grating, respectively. A closed-cycle helium cryostat (Cryo Industries) was used for temperature-dependent measurements. A cryogen-free 7 T superconducting magnet system (Cryo Industries) was interfaced with our cryostat for magneto-Raman measurements. All thermal cycles were performed at a base pressure lower than 7$\times$10$^{–7}$ Torr.

The high-field photoluminescence measurements were performed in a 14T Quantum Design Physical Properties Measurement System equipped with a custom probe containing a piezo stack (attocube) that translates the sample under the objective (attocube, NA = 0.65). The excitation was provided by a 532 nm laser and the spectra were recorded using an IsoPlane 320 (Princeton Instruments) with a TEC CCD camera.\\

\begin{center}
    \textbf{APPENDIX B: Raman Tensor and Selection Rule}
\end{center}

\begin{figure}[ht!]
\includegraphics[width=0.49\textwidth]{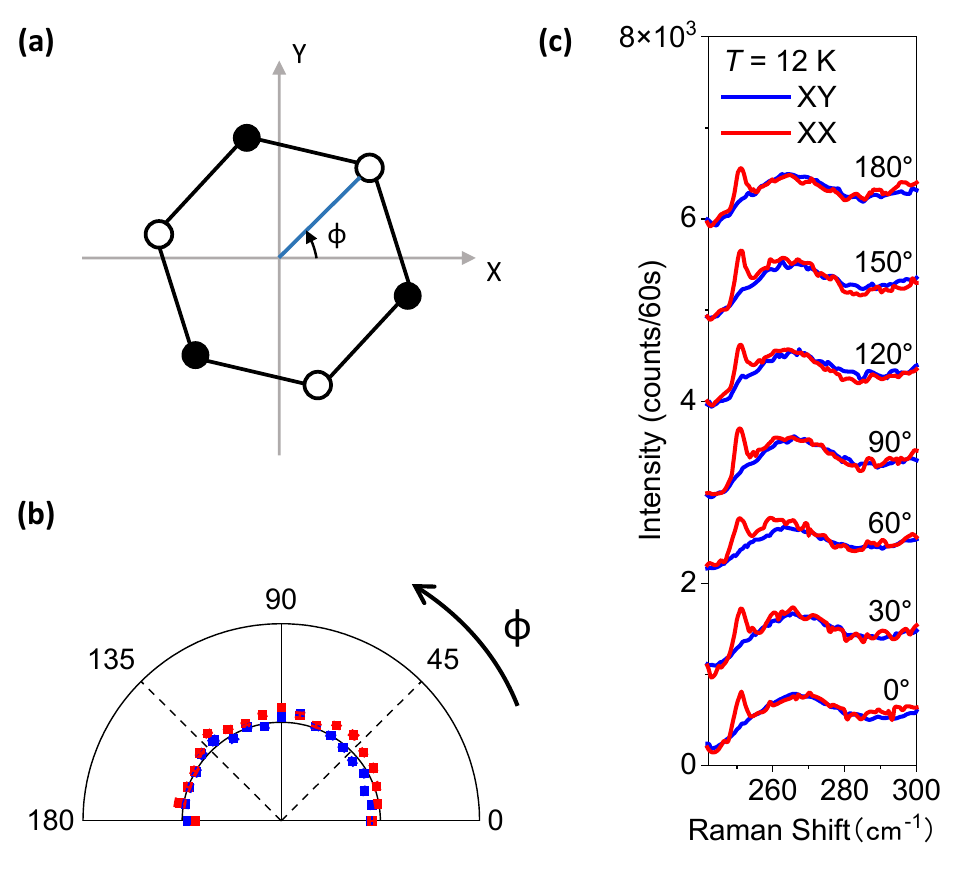}
\caption{\label{Rotation} \justifying\small$\textbf{(a)}$ Schematic of the in-plane rotation of monolayer $\ce{MoS2}$ sample with an azimuthal angle $\phi$. $\textbf{(b)}$ Raman intensity of the 270 cm$^{-1}$ mode in the XX (red) and XY (blue) channels as a function of $\phi$ in a polar plot.  $\textbf{(c)}$ Raman spectra in the XX and XY channels at selected angles of $\phi$. The sharp peak at $\sim$250 cm$^{-1}$ appears only in the XX channel. It is an $A$ phonon mode that is not detectable in the $\sigma^{+}\sigma^{-}/\sigma^{-}\sigma^{+}$ channels. The spectra are vertically offset for clarity.}
\end{figure}

The intensity of Raman scattering is defined as $I = |\langle e_{s} |\textbf{\emph{R}}|e_{i}\rangle|^{2}$, where $e_s$ and $e_i$ denote the polarization vectors of scattered and incident photons and $\textbf{\emph{R}}$ indicates the Raman tensor. In the backscattering geometry, the left- and right-handed circularly polarized light can be assigned as
$\mathrm{\sigma^{+} = \begin{pmatrix}
1  \\
i 
\end{pmatrix}}$
and 
$\mathrm{\sigma^{-} = \begin{pmatrix}
1  \\
-i 
\end{pmatrix}}$, respectively. For linear polarization channels, the horizontal and vertical polarization can be denoted as $\mathrm{X = \begin{pmatrix}
1  \\
0 
\end{pmatrix}}$
and 
$\mathrm{Y = \begin{pmatrix}
0  \\
1 
\end{pmatrix}}$, respectively. \\

In the off-resonance condition, the doubly degenerate $E^{''}$ mode of monolayer \ce{MoS2} (point group $D_{3h}$) has a Raman tensor in the form of $\begin{pmatrix}
0 & 0 & 0\\
0 & 0 & d\\
0 & d & 0
\end{pmatrix}$ and $\begin{pmatrix}
0 & 0 & -d\\
0 & 0 & 0\\
-d & 0 & 0
\end{pmatrix}$ ~\cite{Zhang2015PhononMaterial}. They are Raman active, but cannot be detected in the backscattering geometry.\\

Resonant excitation at 633 nm changes the selection rules. To derive the Raman tensor for the new mode at 270 cm$^{-1}$ in the backscattering geometry, we start with the most general $2\times2$ matrix $\mathrm{\textbf{\emph{M}} = \begin{pmatrix}
a & b \\
c & d 
\end{pmatrix}}$. In the circular polarization channels, the Raman intensity in the $\sigma^{+}\sigma^{+}/\sigma^{-}\sigma^{-}$ channels can be written as follows:

\begin{center}
$|\langle \sigma^{+} |\textbf{\emph{M}}|\sigma^{+}\rangle|^{2} = |(a+d) + (b-c)i|^2$

$|\langle \sigma^{-} |\textbf{\emph{M}}|\sigma^{-}\rangle|^{2} = |(a+d) - (b-c)i|^2$\\
\end{center}

\noindent Because the new mode is absent in the $\sigma^{+}\sigma^{+}/\sigma^{-}\sigma^{-}$ channels, we obtain $d=-a$ and $b=c$. Thus we have the Raman tensor in the form of $\mathrm{\textbf{\emph{R}} = \begin{pmatrix}
a & c \\
c & -a 
\end{pmatrix}}$, which indeed has the symmetry of a doubly degenerate $E$ mode.\\

We then examine the selection rule in the linear polarization channels by rotating the sample from 0$^\circ$ to 180$^\circ$ (see Fig. ~\ref{Rotation}(a)). The Raman intensity in the linearly parallel (XX) and crossed (XY) polarization channels can be written as:

\begin{center}
$|\langle X |\textbf{\emph{R}}|X\rangle|^{2} = (a \mathrm{cos}2\phi-c \mathrm{sin}2\phi)^2$

$|\langle X |\textbf{\emph{R}}|Y\rangle|^{2} = (a \mathrm{sin}2\phi+c \mathrm{cos}2\phi)^2$\\
\end{center}

As shown in Fig.~\ref{Rotation}(b)\&~\ref{Rotation}(c), the new mode at 270 cm$^{-1}$ has identical Raman intensity in the XX and XY channels without angular anisotropy, that is, $|\langle X |\textbf{\emph{R}}|X\rangle|^{2} = |\langle X |\textbf{\emph{R}}|Y\rangle|^{2}$ = Constant. We thus obtain $c=ai$ and the Raman tensor becomes $\mathrm{\textbf{\emph{R}} = a\begin{pmatrix}
1 & i \\
i & -1 
\end{pmatrix}}$. If \textbf{\textit{R}} represents the left-handed phonon mode ($\Omega_{+}$), then its right-handed counterpart ($\Omega_{-}$) must be $\mathrm{\textbf{\emph{R'}} = a\begin{pmatrix}
1 & -i \\
-i & -1 
\end{pmatrix}}$, so that they are connected by a mirror operation. Based on $\textbf{\emph{R}}$ and $\mathrm{\textbf{\emph{R'}}}$, the helicity selection rule for the $\Omega_{\pm}$ mode observed in the resonant condition is summarized in \textbf{TABLE II}, where the $\Omega_{+}$($\Omega_{-}$) mode can be exclusively selected by the 
$\sigma^{+}\sigma^{-}(\sigma^{-}\sigma^{+})$ channel.\\

\begin{table}[ht!]
\textbf{\caption{Helicity selection rule for the $\Omega_{\pm}$ mode.}}
\vspace{5pt}
\centering
\renewcommand{\arraystretch}{1}
\begin{tabular}{|c|c|c|c|c|c|c|}

\hline
\textbf{Raman tensor} & $\sigma^{+} \sigma^{+}$ & $\sigma^{-} \sigma^{-}$ & $\sigma^{+} \sigma^{-}$ & $\sigma^{-} \sigma^{+}$ \\ \hline

$R(\Omega_{+}$)$ = a\begin{pmatrix}
1 & i \\
i & -1 
\end{pmatrix}$  &  0  &   0   & $\mathrm{16|a|^2}$     & 0 \\
\hline
$R'(\Omega_{-}$)$ = a\begin{pmatrix}
1 & -i \\
-i & -1 
\end{pmatrix}$  &  0  &   0   &  0    & $\mathrm{16|a|^2}$ \\
\hline
\end{tabular}
\end{table}
\label{table:Tabel2}

\begin{center}
    \textbf{APPENDIX C: Methods of Theoretical Calculations}
\end{center}

We first use density functional theory calculations to produce phonon frequencies and vibration modes. The spring constants and dynamic matrix are obtained in this process for subsequent model calculations. We consider the unit cell of the monolayer $\ce{MoS2}$, which consists of 3 atoms. The spring force constants are obtained by the finite difference method within the generalized gradient approximation (GGA). Using the VASP package, the spring force constants are computed by $7\times7\times1$ supercells and a $3\times3\times1$ GAMMA grid in the finite displacement method, which is implemented using PHONOPY, an open-source software package. We use an energy cut-off of 650 eV after the convergence tests. In the lattice relaxation process, the convergence thresholds of the total energy and Hellmann-Feynman force are set as $10^{-8}$ eV/$\mathrm{\AA}$ and $10^{-8}$ eV/$\mathrm{\AA}$. The optimized lattice constant a = 3.18 $\mathrm{\AA}$ is obtained and the vacuum layer is fixed at 20 $\mathrm{\AA}$. The resulting 6 modes can be described by the irreducible representation, $E^{''}+2E^{'}+ 2A_{2}^{''}+A_{1}^{'}$.

To determine the Raman spectra, we calculated the derivative of the dielectric susceptibility with respect to the phonon displacement by the Feynman-Hellman method. The corresponding Raman tensors are:
$\mathrm{\Re _{E^{''}}^{\mu}(1):\begin{bmatrix}
-0.936 & 0.360 \\ 
 0.360 & 0.936 
\end{bmatrix}}$ and $\mathrm{\Re _{E^{''}}^{\mu}(2):\begin{bmatrix}
-0.191 & -0.997 \\ 
 -0.997 & 0.191 
\end{bmatrix}}$ for the $E^{''}$ mode (279.2 cm$^{-1}$).

In the presence of an external static magnetic field (\textbf{\textit{H}}), the effective spring constant ($\varphi$) can be expanded in the components of the magnetic field:

\begin{equation}
\varphi_{\alpha\beta}(lk;l'k')= \varphi^{(0)}_{\alpha\beta}(lk;l'k')+i\sum_{\gamma}\varphi^{(1)}_{\alpha\beta\gamma}(lk;l'k')H_{\gamma}
\end{equation}

\noindent where $\alpha$, $\beta$, $\gamma$ label the Cartesian coordinates, $l$ and $l'$ are the primitive unit cells of the crystal, and $k$ and $k'$ are the atoms in a primitive unit cell. The frequency of the Brillouin zone center mode can be obtained from the solutions to the eigenvalue equation

\begin{equation}
\omega^2_{sj}\epsilon_{\alpha}(k;sj\lambda)=\sum_{k'\beta}D_{\alpha\beta}(kk';\textbf{\emph{H}})\epsilon_{\beta}(k';sj\lambda)
\end{equation}

\noindent where the dynamic matrix $D_{\alpha\beta}(kk';\textbf{\emph{H}})$ is defined as

\begin{equation}
D_{\alpha\beta}(kk';\textbf{\emph{H}})=\frac{1}{\sqrt{M_{k}M_{k}}}\sum_{l'}\varphi_{\alpha\beta}(lk;l'k')
\end{equation}

\noindent and $M_k$ is the mass of the $k^\mathrm{th}$ atom, $s$ labels the irreducible representation of the phonon mode, $j$ denotes the modes of different frequencies that belong to the same representation, and $\lambda$ is the dimension of the representation.

For doubly degenerate phonon mode $E^{''}$ with frequency of $\omega_j$ and $\textbf{\emph{H}}=H e_z$, the normalized polarization vector $\Tilde{\epsilon}_{j\lambda}(\lambda=1,2)$ can be written as 
$\Tilde{\epsilon}_{j\lambda} = \Sigma_{\lambda'}C_{\lambda\lambda'}\epsilon_{j\lambda'}$, where $\epsilon_{j\lambda}$ satisfies $D^{(0)}\epsilon_{j\lambda} = \omega^2_j\epsilon_{j\lambda}$. Using Eq. (6), one can obtain the new frequency $\omega_{jH}$ from the eigenvalue equation

\begin{gather}
 \begin{vmatrix} \omega^2_{jH} - \omega^2_{j} & iK_{12}H \\ iK_{21}H & \omega^2_{jH} - \omega^2_{j} \end{vmatrix}=0
\end{gather}

\noindent Since $K_{\lambda\lambda'}=\epsilon^{\dagger}_{j\lambda}D^{(1)}\epsilon_{j\lambda'}$ is an antisymmetric matrix, i.e. $k_{12}=-K_{21}=K$ ~\cite{schaack1976observation, schaack1977magnetic}. By solving Eq. (7), we obtain the linear Zeeman splitting of the $E^{''}$ mode and the circularly polarized eigenvectors.

\section{Acknowledgments}

We acknowledge the helpful discussion with Roberto Merlin, Dali Sun, Yafei Ren, Xiaowei Zhang, and Zhengguang Lu. W.J. acknowledges support by NSF EPM Grant No. DMR-2129879 and Auburn University Intramural Grants Program. C.T. acknowledges financial support from the Alabama Graduate Research Scholars Program (GRSP) funded through the Alabama Commission for Higher Education and administered by the Alabama EPSCoR. G.Y., C.N., and R.H. are supported by NSF Grants No. DMR-2104036 and No. DMR-2300640. T.W. is sponsored by the Postgraduate Research $\&$ Practice Innovation Program of Jiangsu Province (Grant No. KYCX23$\_$1677). L.X. and D.S. acknowledge support from the U.S. Department of Energy (DE-FG02-07ER46451) for high-field magneto-photoluminescence measurements performed at the National High Magnetic Field Laboratory, which is supported by the National Science Foundation through NSF Grant No. DMR-1644779 and the state of Florida.

\bibliographystyle{apsrev4-1}
\nocite{apsrev41Control}
\bibliography{references.bib}

\end{document}


\preprint{}

\title{Supplementary Material \\ \ \\ 
\normalsize Exciton-activated effective phonon magnetic moment in monolayer \ce{MoS2}}

\author{Chunli Tang}
\thanks{These two authors contributed equally}
\affiliation{Department of Electrical and Computer Engineering, Auburn University, Auburn, Alabama 36849, USA}
\affiliation{Department of Physics, Auburn University, Auburn, Alabama 36849, USA}

\author{Gaihua Ye}
\thanks{These two authors contributed equally}
\affiliation{Department of Electrical and Computer Engineering, Texas Tech University, Lubbock, Texas 79409, USA}

\author{Cynthia Nnokwe}
\affiliation{Department of Electrical and Computer Engineering, Texas Tech University, Lubbock, Texas 79409, USA}

\author{Mengqi Fang}
\affiliation{Department of Mechanical Engineering, Stevens Institute of Technology, Hoboken, New Jersey 07030, USA}

\author{Li Xiang}
\affiliation{National High Magnetic Field Laboratory, Tallahassee, Florida 32310, USA}

\author{Masoud Mahjouri-Samani}
\affiliation{Department of Electrical and Computer Engineering, Auburn University, Auburn, Alabama 36849, USA}

\author{Dmitry Smirnov}
\affiliation{National High Magnetic Field Laboratory, Tallahassee, Florida 32310, USA}

\author{Eui-Hyeok Yang}
\affiliation{Department of Mechanical Engineering, Stevens Institute of Technology, Hoboken, New Jersey 07030, USA}

\author{Tingting Wang}
\affiliation{Phonon Engineering Research Center of Jiangsu, School of Physics and Technology, Nanjing Normal University, Nanjing 210023, China}

\author{Lifa Zhang}
\affiliation{Phonon Engineering Research Center of Jiangsu, School of Physics and Technology, Nanjing Normal University, Nanjing 210023, China}

\author{Rui He}
\email{rui.he@ttu.edu}
\affiliation{Department of Electrical and Computer Engineering, Texas Tech University, Lubbock, Texas 79409, USA}

\author{Wencan Jin}
\email{wjin@auburn.edu}
\affiliation{Department of Physics, Auburn University, Auburn, Alabama 36849, USA}
\affiliation{Department of Electrical and Computer Engineering, Auburn University, Auburn, Alabama 36849, USA}

\maketitle
\tableofcontents

\newpage
\section{$\textbf{S1.}$ Atomic force microscope characterization of CVD-grown monolayer $\ce{MoS2}$}
\begin{figure*}[ht!]
\includegraphics[width=1.0\textwidth]{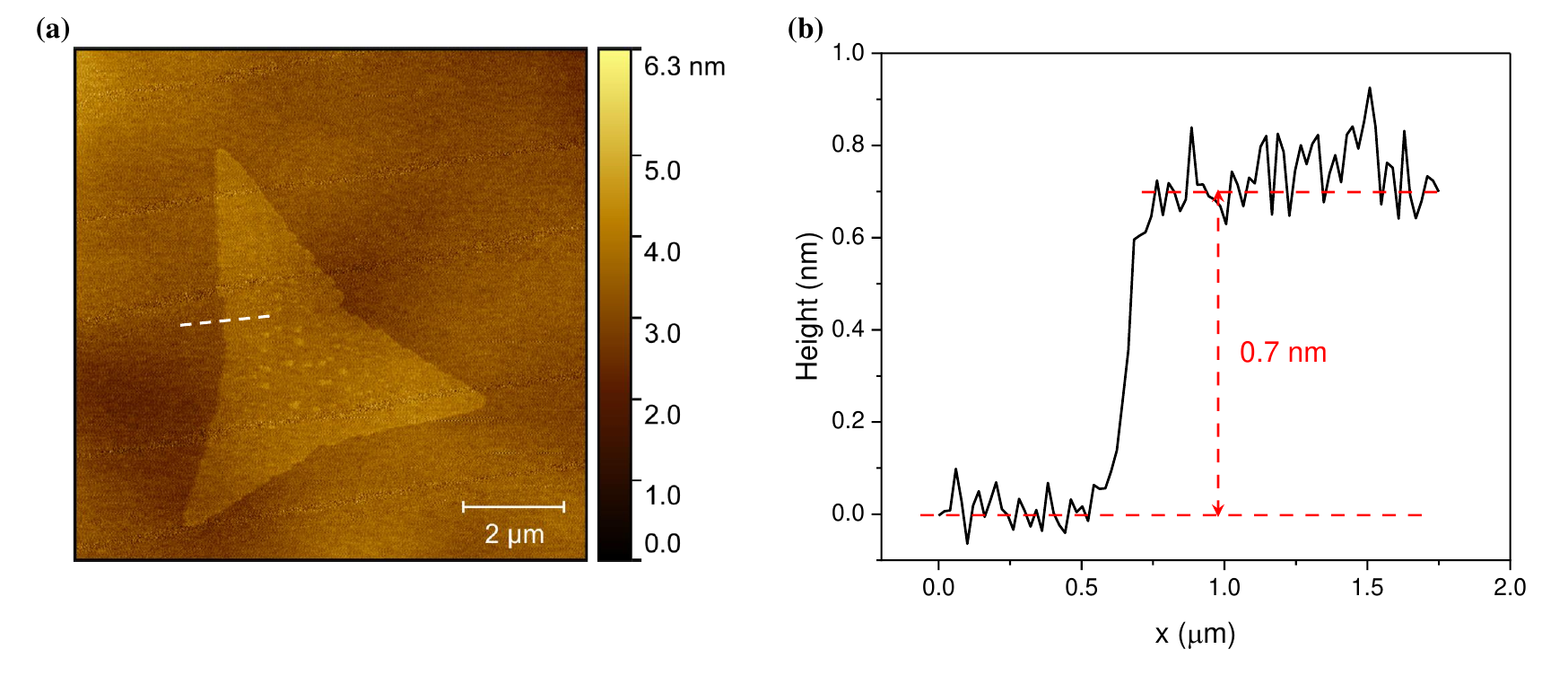}
\caption{\label{AFM} \justifying \small $\textbf{(a)}$ AFM image of a triangle CVD-grown $\ce{MoS2}$ flake on the $\ce{SiO_{2}/Si}$ substrate. The scale bar is 2 $\mu$m. $\textbf{(b)}$ An AFM line profile across the edge of the flake as marked by the white dashed line in (a).}
\end{figure*}

Figure \ref{AFM} shows the characterization of a representative CVD-grown $\ce{MoS2}$ flake. The flake has a uniform surface as shown in the atomic force microscope (AFM) image in Fig. \ref{AFM}(a). The height profile across the edge of the flake confirms its monolayer thickness. 

\newpage
\section{$\textbf{S2.}$ Comparison with resonant Raman spectrum of $\ce{MoS2}$ in the literature}
\begin{figure*}[ht!]
\includegraphics[width=1.0\textwidth]{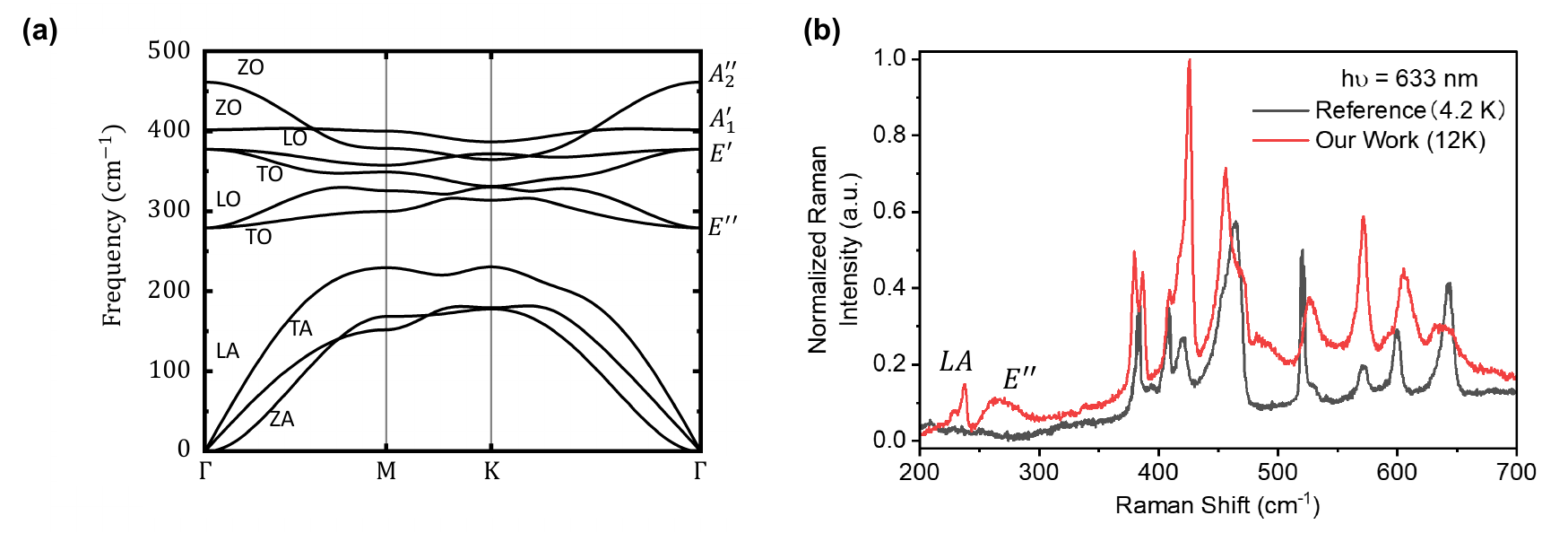}
\caption{\label{Raman Resonant} \justifying\small $\textbf{(a)}$ Calculated phonon dispersion of monolayer $\ce{MoS2}$.  $\textbf{(b)}$ Raman spectra acquired using 633 nm excitation from Ref.~\cite{10.1063/1.4867502} (black) and our work (red).}
\end{figure*}

In Fig.~\ref{Raman Resonant}(b), we compare the Raman modes in our spectrum with the one from the literature ~\cite{10.1063/1.4867502}. Both spectra are acquired using 633 nm excitation at low temperature. In the spectral range of 300-700 cm$^{-1}$, the Raman peaks in our data match the reference well. In the range of 200-300 cm$^{-1}$, we found two new modes in our spectrum. Compared with the calculated phonon band shown in Fig.~\ref{Raman Resonant}(a), the sharp doublet at $\sim$240 cm$^{-1}$ can be assigned to the LA mode at M or K point~\cite{bhatnagar2022temperature}, and the broad peak centered at $\sim$270 cm$^{-1}$ matches the frequency of the $E^{''}$ mode. 

\newpage
\section{$\textbf{S3.}$ Magnetic-field dependent helicity-resolved Raman data of $E^{'}$ and $A_{1}^{'}$ modes}
\begin{figure*}[ht!]
\includegraphics[width=0.45\textwidth]{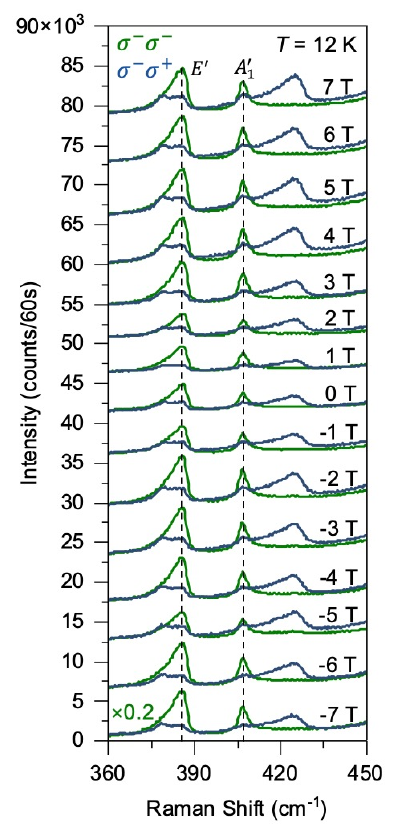}
\caption{\label{EandAmode} \small \justifying Helicity-resolved Raman spectra showing the $E^{'}$ and $A_{1}^{'}$ phonon modes at varying magnetic field at 12 K. The spectra are vertically offset for clarity and the spectra in the $\sigma^{-} \sigma^{-}$ channel are scaled by a factor of 0.2.}
\end{figure*}

Figure~\ref{EandAmode} shows the magnetic field dependent measurements of the $E^{'}$ and $A_{1}^{'}$ phonon modes. As marked by the dashed lines, in both helicity-reserved ($\sigma^{-} \sigma^{-}$) and helicity-changed ($\sigma^{-} \sigma^{+}$) channels, the frequency of $E^{'}$ and $A_{1}^{'}$ is magnetic field independent. 

\newpage
\section{$\textbf{S4.}$ Helicity-resolved Raman scattering data of exfoliated monolayer $\ce{MoS2}$}

\begin{figure*}[ht!]
\includegraphics[width=0.64\textwidth]{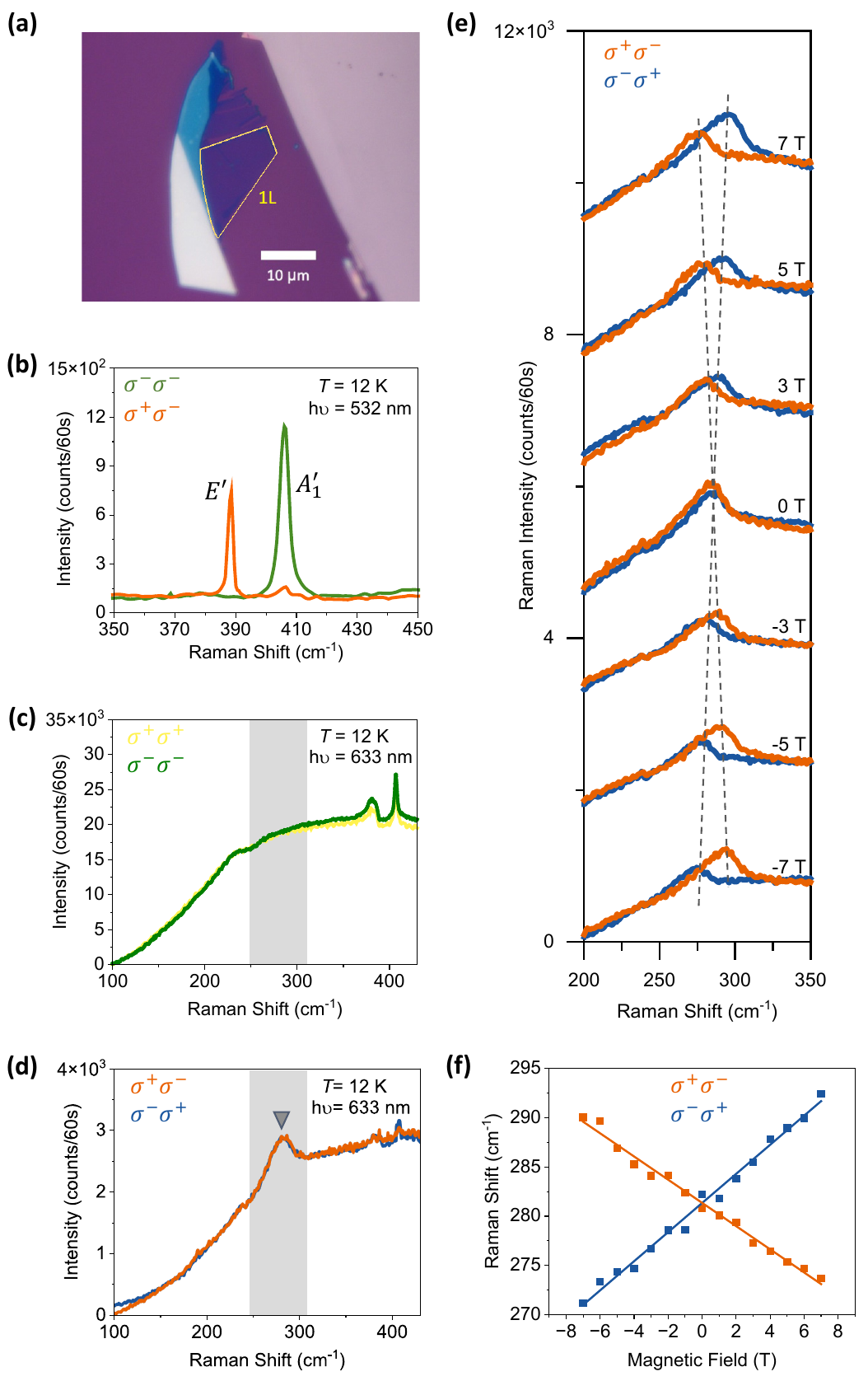}
\caption{\label{Exfoliated}\small \justifying $\textbf{(a)}$ Optical microscope image of the exfoliated \ce{MoS2} flake. The yellow box encloses the monolayer (1L) region. $\textbf{(b)}$ In the off-resonance condition, $E^{'}$ and $A_{1}^{'}$ Raman modes acquired using 532 nm laser. In the on-resonance condition using 633 nm laser, the helicity selection rule of the $E^{''}$ mode is examined in $\textbf{(c)}$ $\sigma^+ \sigma^+$/$\sigma^- \sigma^-$ and $\textbf{(d)}$ $\sigma^+ \sigma^-$/$\sigma^- \sigma^+$ polarization channels. \textbf{(e)} Magnetic field dependent Raman spectra in the $\sigma^+ \sigma^-$/$\sigma^- \sigma^+$ polarization channels. The grey dashed lines are the guide to the eye of the phonon Zeeman splitting. \textbf{(f)} Split chiral mode frequencies as a function of the magnetic field with linear fits. All the Raman spectra are measured at 12 K.}
\end{figure*}

We prepared a pristine exfoliated monolayer \ce{MoS2} sample (see optical microscope image in Fig.~\ref{Exfoliated}(a)) and measured the helicity-resolved Raman spectroscopy at 12 K. In the off-resonance condition at 532 nm (see Fig.~\ref{Exfoliated}(b)), the energy separation between the $E^{'}$ and $A_{1}^{'}$ phonon modes is $\sim$18 cm$^{-1}$, which is consistent with the exfoliated monolayer \ce{MoS2}. In the on-resonance condition at 633 nm, as shown in Fig.~\ref{Exfoliated}(c) \&~\ref{Exfoliated}(d), the broad $E^{''}$ mode appears in the helicity-changed $\sigma^+ \sigma^-$/$\sigma^- \sigma^+$ channels and is absent in the helicity-reserved $\sigma^+ \sigma^+$/$\sigma^- \sigma^-$ channels. Therefore, we confirm the $E^{''}$ mode exists in both CVD and exfoliated \ce{MoS2} samples following the same helicity selection rule. Note that in both circularly parallel and crossed channels, the Raman spectra of the exfoliated sample have a profound background from the tail of the photoluminescence.

Figure~\ref{Exfoliated}(e) shows the magnetic field dependent measurements of the exfoliated sample. The lineshapes of the chiral mode become skewed due to the profound background. Therefore, the frequency of the chiral mode is determined using the second-derivative method. As shown in Fig.~\ref{Exfoliated}(f), the frequency of the split modes exhibits a linear response to the magnetic field. Linear fits to the experimental data yield slopes of $\mathrm{1.18\pm0.04}$ cm$^{-1}$/T ($\sigma^+ \sigma^-$ channel) and $\mathrm{1.47\pm0.04}$ cm$^{-1}$/T ($\sigma^- \sigma^+$ channel), respectively, which are similar to the values we obtained from the CVD samples. Note that at 0 T, the doubly degenerate $E^{''}$ mode of the exfoliated \ce{MoS2} is located at the DFT-predicted position of 280 cm$^{-1}$. The deviation of the peak position in CVD samples (270 cm$^{-1}$) implies that this mode is sensitive to the exciton and specific band structure of \ce{MoS2}.

\clearpage

\newpage
\section{$\textbf{S5.}$ Power-dependent measurements of the split chiral modes at 7 T}

\begin{figure*}[ht!]
\includegraphics[width=0.95\textwidth]{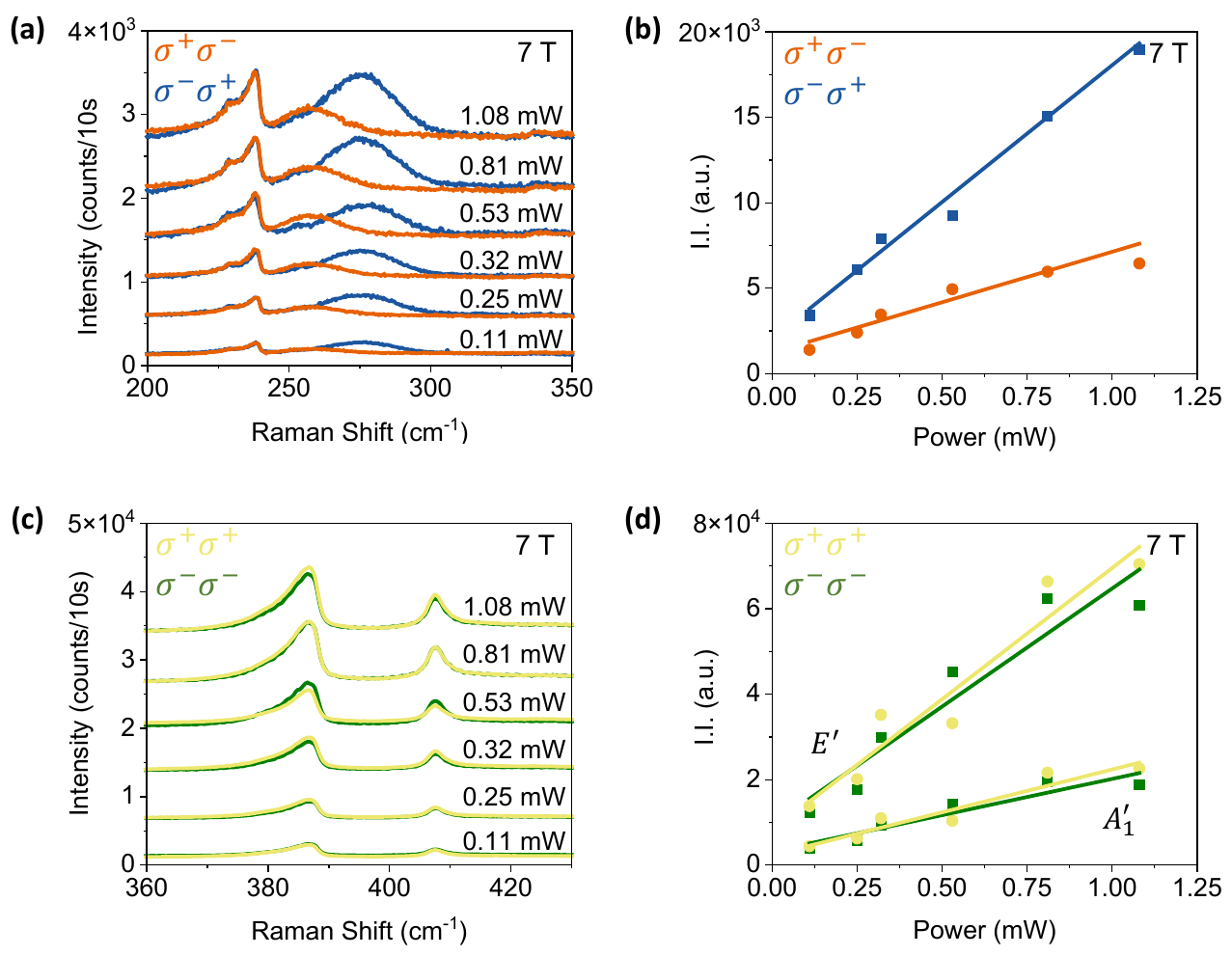}
\caption{\label{power}\small \justifying $\textbf{(a)}$ Raman spectra in the $\sigma^+ \sigma^-$/$\sigma^- \sigma^+$ channels acquired using varying laser powers. $\textbf{(b)}$ Integrated intensity of the split modes as a function of laser power. $\textbf{(c)}$ Raman spectra in the $\sigma^+ \sigma^+$/$\sigma^- \sigma^-$ channels acquired using varying laser powers. $\textbf{(d)}$ Integrated intensity of the $E^{'}$ and $A_{1}^{'}$ modes as a function of laser power. All the Raman spectra are measured at 12 K and 7 T.}
\end{figure*}

We carry out power-dependent measurements of the split modes at 7 T. As described in APPENDIX A, the data shown in the main text are collected at 0.5 mW. Figure ~\ref{power}(a) shows the Raman spectra in $\sigma^+ \sigma^-$/$\sigma^- \sigma^+$ channels acquired from 0.11 mW to 1.08 mW, and no laser heating is observed in this process. The integrated intensities (I.I.) of the split modes linearly scale up with the increasing power (see Fig. ~\ref{power}(b)), showing the same behavior as the $E^{'}$ and $A_{1}^{'}$ phonon modes (see Fig. ~\ref{power}(c) \& ~\ref{power}(d)).

\newpage
\section{$\textbf{S6.}$ Temperature-dependent measurements of exfoliated monolayer \ce{MoS2}}

\begin{figure}[h!]
\includegraphics[width=0.75\textwidth]{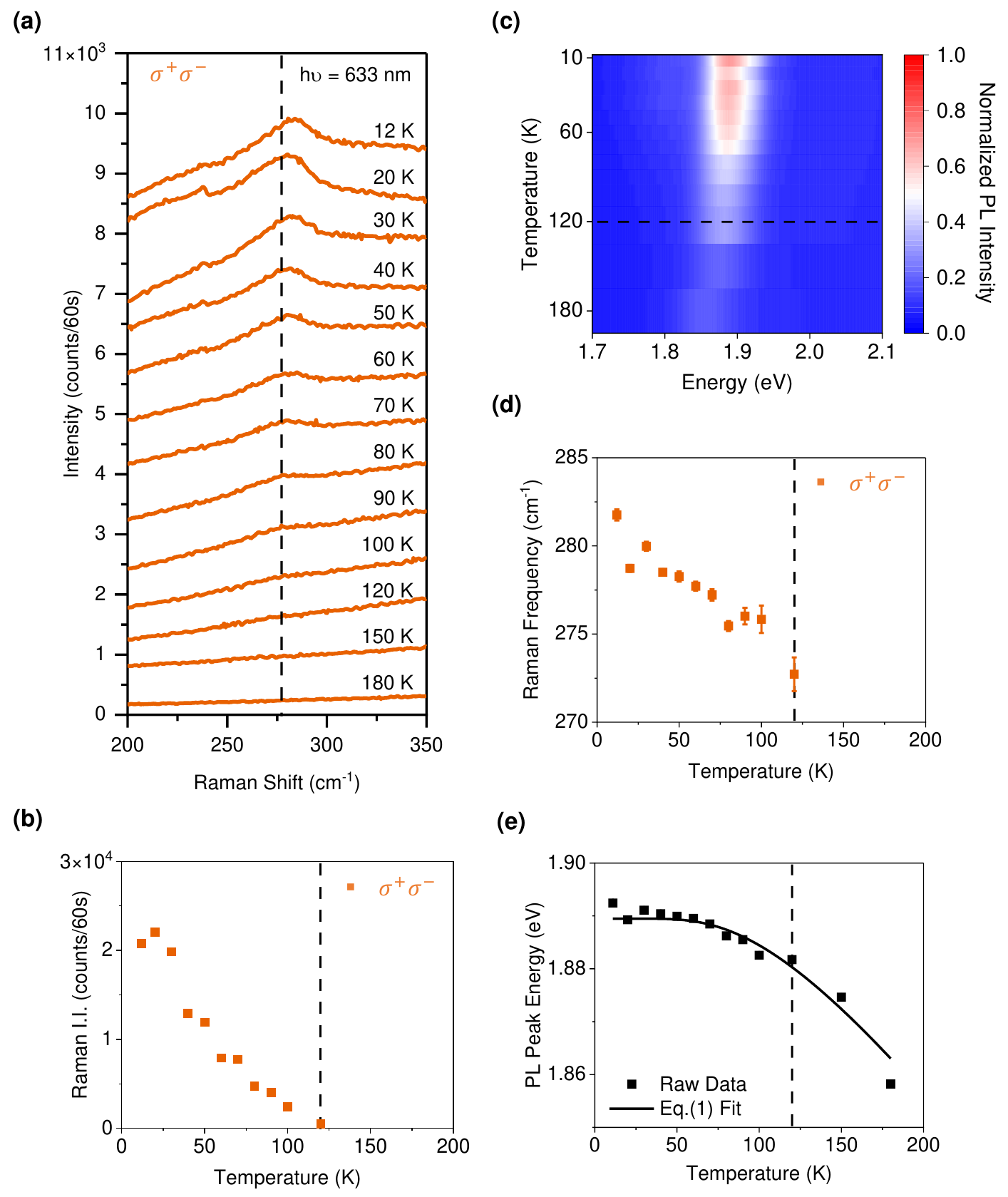}
\caption{\label{T_exf} \small \justifying \textbf{(a)} Temperature-dependent Raman spectra in the $\sigma^+ \sigma^-$ polarization channel at the excitation wavelength of 633 nm. The spectra are vertically offset for clarity. \textbf{(b)} Integrated intensity (I.I.) of the chiral phonon mode as a function of temperature. \textbf{(c)} Heatmap of the temperature-dependent PL spectra of $A$ exciton measured using 532 nm excitation laser. The colorscale represents the normalized PL intensity. \textbf{(d)} Raman frequency of the $\Omega_{\pm}$ mode and \textbf{(e)} PL peak energy of the $A$ exciton as a function of temperature. The solid black curve in (e) is the fit to Eq. (1) in the main text. The black dashed lines in panels (b)-(e) mark the temperature of 120 K, where the chiral phonon mode emerges and the PL intensities of $A$ exciton rapidly grow.}
\end{figure}

\newpage
\section{$\textbf{S7.}$ Temperature-dependent measurements of PL at 633 nm excitation}
\begin{figure*}[ht!]
\includegraphics[width=0.95\textwidth]{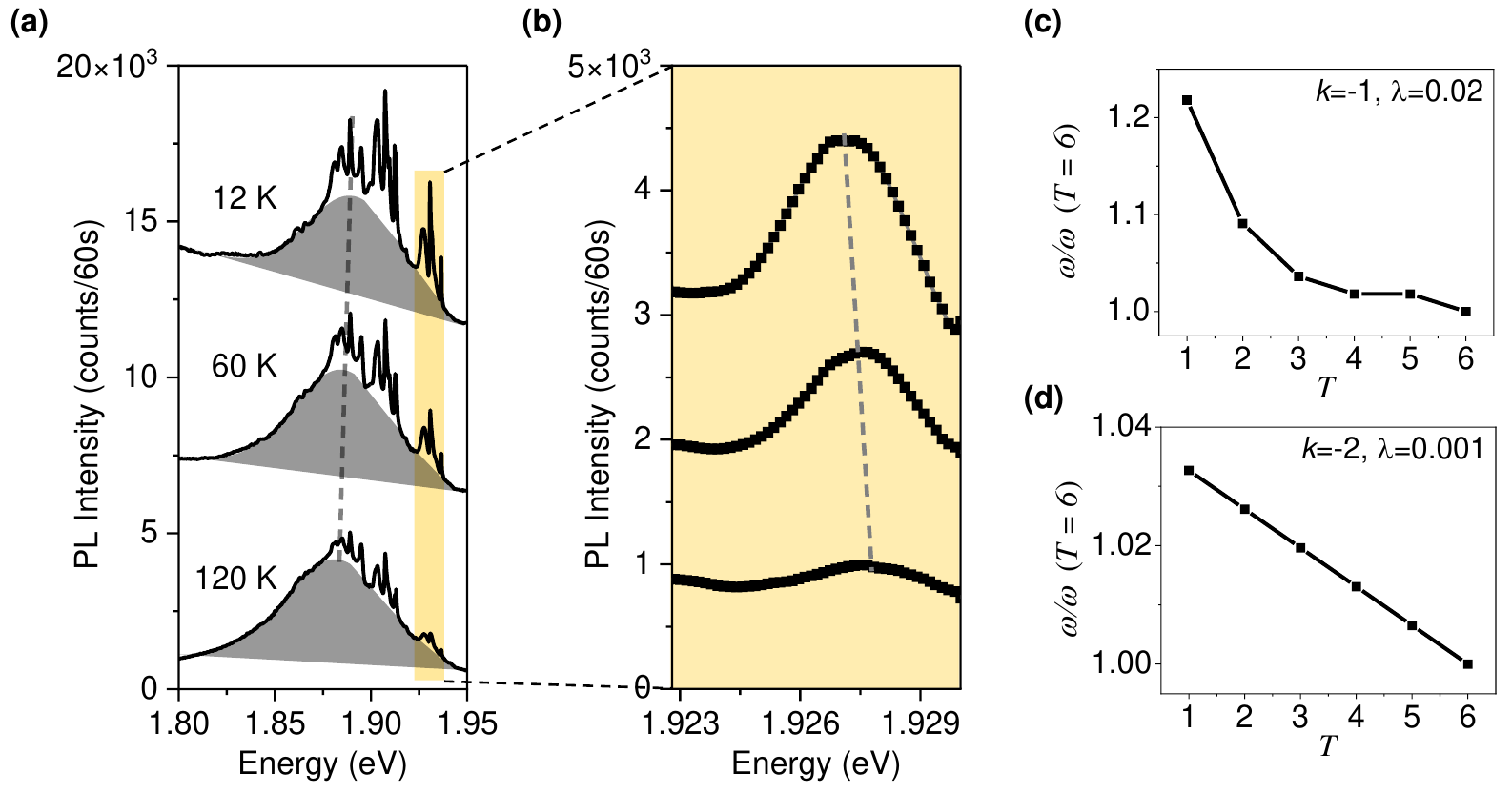}
\caption{\label{PL633nm} \small \justifying \textbf{(a)} PL spectra acquired using 633 nm excitation laser at varying temperatures. The spectra are vertically offset for clarity. The grey broad profiles are fits to the $A$ exciton. \textbf{(b)} Zoom in of the yellow shaded block in (a). The grey dashed lines are the guide to the eye. \textbf{(c) \& (d)} The simulated phonon frequency as a function of temperature ($T$) for representative $k$ and $\lambda$ values. The temperature is in dimensionless unit and the phonon frequency ($\omega$) is normalized to the value at $T=6$.}
\end{figure*}

We carry out temperature-dependent PL measurements of CVD \ce{MoS2} using the 633 nm excitation laser, in which we can track both $A$ exciton and the 270 cm$^{-1}$ Raman mode simultaneously in their absolute photon energies. Going from 120 K to 12 K, the $A$ exciton shifts towards higher energy by $\sim$11 meV (see Fig. ~\ref{PL633nm}(a)), which agrees with the value we obtain from the 532 nm laser (see Fig. 3(e) in the main text). For the 270 cm$^{-1}$ Raman mode, as we zoom in the energy range around 1.927 eV (see Fig. ~\ref{PL633nm}(b)), the absolute photon energy shifts towards lower energy by $\sim$0.5 meV ($\sim$4 cm$^{-1}$), which is consistent with the shift of this mode resolved in Raman spectra. In absolute photon energy, the $A$ exciton and the 270 cm$^{-1}$ Raman mode shift towards opposite directions, confirming their distinct origins.

Upon cooling, the chiral phonon mode shifts by $\sim$2\% without saturation at the lowest temperature, indicating the complicated nature of this mode beyond harmonic approximation. To evaluate the anharmonicity of this mode, we carry out theoretical calculations based on the phonon Hamiltonian below
\begin{eqnarray}\label{eq:hamiltonian}
    \hat{H}_{ph} &= &\frac{\hat{P}^2}{2m}+\frac{1}{2}k\hat{x}^2 + \frac{1}{4}\lambda \hat{x}^4
\end{eqnarray}
\noindent where $k$ is the spring constant, $m$ is the effective mass, $\hat{P}$ is the momentum, and $\lambda$ is the anharmonic strength. Using the formulation of Ref.~\cite{PhysRevB.79.214306}, we calculate the temperature dependence of optical conductivity defined as
\begin{equation}
\frac{\sigma(\omega)}{\sigma_0} = i\omega \sum_{\omega_{mn}>0}|\langle n|(\frac{x}{x_0})|m \rangle|^2\frac{e^{-\beta E_m}-e^{-\beta E_n}}{Z}(\frac{1}{\omega-\omega_{nm}+i\Gamma_0/2}-\frac{1}{\omega+\omega_{nm}+i\Gamma_0/2})
\end{equation}
\noindent where $E_n$, $|n\rangle$ are the energies and eigenmodes of $\hat{H}_{ph}$ which we compute numerically using perturbation theory. Here we consider phonon lifetime $\Gamma_0$ as a constant, while as demonstrated in Ref.~\cite{PhysRevLett.128.075901}, the model remains valid for a temperature-dependent phonon lifetime.

As shown in Fig.~\ref{PL633nm}(c)\&~\ref{PL633nm}(d), for finite $\lambda$, a rapid phonon frequency shift is obtained in good agreement with the trend of our experimental data. Similar results were also reported in SnSe ~\cite{PhysRevB.98.224309}. Note that our simulation indicates that the anharmonic strength of the chiral phonon mode in \ce{MoS2} is relatively weak ($\lambda/k$ is approximately at the scale of $10^{-3}$). In contrast, the anharmonicity of chiral TO phonon in PbTe ~\cite{PhysRevLett.128.075901} is very strong, resulting in softening phonon as temperature decreases.

\clearpage

\newpage
\bibliographystyle{apsrev4-1}
\nocite{apsrev41Control}
\bibliography{Supplement.bib}